\documentclass{article}
\usepackage{spconf,amsmath,graphicx}
\usepackage{multirow,booktabs}
\usepackage{multicol}
\usepackage{mathabx}

\usepackage{enumitem}
\usepackage{url}
\usepackage{hyperref} 
\usepackage{pdfpages}

\title{P\MakeLowercase{er}SIM: Multi-Resolution Image Quality Assessment in the Perceptually Uniform Color Domain }
\name{Dogancan Temel and Ghassan AlRegib}

\address{Center for Signal and Information Processing (CSIP)\\
School of Electrical and Computer Engineering\\
Georgia Institute of Technology, Atlanta, GA, 30332-0250 USA\\
\{cantemel,alregib\}@gatech.edu}

\begin{document}
%\ninept

\onecolumn % make sure you keep this coverpage as one column. In this template, we force the coverpage to be one column with this command and then switch to double column for the remaining of the paper with the \doublecolumn command. 

\begin{description}[labelindent=1cm,leftmargin=3cm,style=multiline]

\item[\textbf{Citation}]{D. Temel and G. AlRegib, "PerSIM: Multi-resolution image quality assessment in the perceptually uniform color domain," 2015 IEEE International Conference on Image Processing (ICIP), Quebec City, QC, 2015, pp. 1682-1686.} \\

\item[\textbf{DOI}]{\url{https://doi.org/10.1109/ICIP.2015.7351087}} \\

\item[\textbf{Review}]{Date added to IEEE Xplore: 10 December 2015} \\

\item[\textbf{Code/Poster}]{\url{https://ghassanalregib.com/publications/}} \\

\item[\textbf{Bib}] {
@INPROCEEDINGS\{Temel2015\_ICIP,\\ 
author=\{D. Temel and G. AlRegib\},\\ 
booktitle=\{2015 IEEE International Conference on Image Processing (ICIP)\},\\ 
title=\{PerSIM: Multi-resolution image quality assessment in the perceptually uniform color domain\},\\ 
year=\{2015\},\\
pages=\{1682-1686\},\\
doi=\{10.1109/ICIP.2015.7351087\},\\ 
month=\{Sept\},\}\\
} \\

\item[\textbf{Copyright}]{\textcopyright 2015 IEEE. Personal use of this material is permitted. Permission from IEEE must be obtained for all other uses, in any current or future media, including reprinting/republishing this material for advertising or promotional purposes,
creating new collective works, for resale or redistribution to servers or lists, or reuse of any copyrighted component
of this work in other works. } \\

\item[\textbf{Contact}]{\href{mailto:alregib@gatech.edu}{alregib@gatech.edu}~~~~~~~\url{https://ghassanalregib.com/}\\ \href{mailto:dcantemel@gmail.com}{dcantemel@gmail.com}~~~~~~~\url{http://cantemel.com/}}
\end{description} 

\thispagestyle{empty}
\newpage
\clearpage

\twocolumn

\maketitle

\vspace{-0.5cm}
\begin{abstract}
\vspace{-0.10cm}
An average observer perceives the world in color instead of black and white. Moreover, the visual system focuses on structures and segments instead of individual pixels. Based on these observations, we propose a full reference objective image quality metric modeling visual system characteristics and chroma similarity in the perceptually uniform color domain (\texttt{Lab}). Laplacian of Gaussian features are obtained in the \texttt{L} channel to model the retinal ganglion cells in human visual system and color similarity is calculated over the \texttt{a} and \texttt{b} channels. In the proposed  perceptual similarity index (\textbf{PerSIM}), a multi-resolution approach is followed to mimic the hierarchical nature of human visual system. LIVE and TID2013 databases are used in the validation and \textbf{PerSIM} outperforms all the compared metrics in the overall databases in terms of ranking, monotonic behavior and linearity. 
 
\end{abstract}

\vspace{-0.3cm}

\section{Introduction}
\label{sec:intro}
\vspace{-0.35cm}

Image quality metrics are designed to estimate the perceived quality.  A full reference objective image quality metric tries to quantify the differences between the original and the distorted image  if both the images are available. Root mean-squared error (RMSE) is calculated by obtaining the pixel-wise difference between the two images, taking the square root of the difference and calculating the mean. RMSE is scaled with the bit depth of the image and mapped using a monotonic logarithmic function to obtain the peak signal-to-noise ratio (PSNR). The authors in \cite{Egiazarian} introduce PSNR-HVS by stretching the contrast block-wise, quantizing the DCT coefficients with the JPEG compression table and removing mean shift. PSNR-HVS is further modified by multiplying the difference between the DCT coefficients with contrast masking metric (PSNR-HVS-M) \cite{Ponomarenko2007}. The authors in \cite{Ponomarenko2011} add mean shift sensitivity and contrast change to the pixel-wise metrics and the modified versions are denoted as PSNR-HA and PSNR-HMA. Signal-to-noise ratio (SNR) is weighted by the authors in  \cite{Mitsa1993} using contrast sensitivity function and the authors in \cite{Chandler2007}  use wavelet-based models of visual masking.

In addition to pixel-wise metrics, structural metrics are also commonly used to estimate the quality of images. The authors in \cite{Wang2004} compare the reference and distorted images in terms of luminance, contrast and structure similarity in the spatial domain to estimate the image quality. These structure-based metrics are also extended to multi-scale (MS-SSIM) \cite{Wang2003}, complex domain (CW-SSIM) \cite{Inter2005} and  information-weighted (IW-SSIM) \cite{Wang2011} versions. Instead of directly using the pixel values, phase and magnitude of the images can also be used separately to estimate the quality. FSIM is a feature similarity index introduced by the authors in \cite{Zhang2011} which consists of phase congruency (PC) and gradient magnitude (GM). PC consists of log-Gabor filter and Gaussian spread function and GM is based on gradient operators. The feature similarity index is further extended as FSIM-c which also includes the color similarity in the YIQ domain. GM is utilized along with the \texttt{LoG} features to obtain joint statistics that can be used for blind image quality assessment by the authors in \cite{Feng2014}. \texttt{LoG} features are also used in \cite{log2014} to directly asses the image quality but they overlook the hierarchical procedure and color perception in the visual system.

\vspace{-0.1cm}

In this paper, we operate in the perceptually uniform \texttt{Lab} color space where luma and chroma information are separated. Retinal ganglion cells in the visual system are modeled using the \texttt{LoG} features in the \texttt{L} channel. Chroma similarity is calculated over the \texttt{a} and \texttt{b} channels. We obtain the similarity maps at different resolutions and calculate the geometric mean of these maps to obtain the multi-resolution similarity maps. \texttt{LoG} and chroma similarities are tuned using the ratios in the 4:2:2 chroma sub-sampling format. After sensitivity tuning, the minimum similarity is selected pixel-wise over the quality maps. Mean pooling is performed over the full map to calculate a single quality value. The resulting value is monotonically mapped by taking the power to obtain the perceptual similarity index \texttt{PerSIM}. 

\vspace{-0.6cm}

\section{PerSIM}
\label{sec:main}
\vspace{-0.4cm}

\subsection{Log Features}
\label{sec:main_Log}
\vspace{-0.2cm}

Instead of using the pixel values as raw data, image features are extracted to represent the images in a more compact and distinctive way. Difference of Gaussian  and Laplacian of Gaussian are among the most commonly used  operators in the image processing literature and the computer vision literature. Difference of Gaussian operators can be used to model the retinal Ganglion cells of the cat as discussed in \cite{Robson1966}. Moreover, the authors in  \cite{Young1987} discuss that Gaussian derivative-like approaches can model neural mechanisms in the human foveal retinal vision. These Gaussian derivative-like approaches also outperform the Gabor filter-based models according to model-free Wiener filter analysis as explained in \cite{Young1987}. In the difference of Gaussian models, standard deviation and the scale of the difference need to be tuned to obtain distinctive features. In case of using various scales, fusion of these models also becomes an issue.  Difference of Gaussian operator can be used as an approximation to the second derivative of Gaussian when the scale is adjusted. And the second derivative of Gaussian corresponds to the Laplacian of Gaussian operator. In order to avoid the tuning of the scale and simplify the problem, we use Laplacian of Gaussian as formulated in Eq. (\ref{eq:LoG}).

\vspace{-0.4cm}

\begin{equation}
\label{eq:LoG}
\hat{LoG}= \frac{1}{\sqrt{2\pi\sigma ^2}}   \frac{m^2+n^2-2\sigma^2}{\sigma^4}  e^{-(m^2+m^2)/(2\sigma^2)} 
\vspace{0.5 cm}
\end{equation}

\vspace{-0.6cm}

The standard deviation of the \texttt{LoG} operator is represented with $\sigma$ and $m$ and $n$ are the respective pixel locations . Reference ($f_1$) and distorted ($f_2$) images are convolved with the \texttt{LoG} operator as formulated in Eq. (\ref{eq:LoG2}) where $i$ corresponds to the image index.

\vspace{-0.3cm}

\begin{equation}
\label{eq:LoG2}
LoG_i=f_i[m,n]\asterisk \hat{LoG}[m,n]
\end{equation}

The similarity between \texttt{LoG} maps is calculated using the familiar similarity formulation that has been part of most of the structural and pixel-wise comparison metrics as expressed in Eq. (\ref{eq:LoG3}).

\vspace{-0.5cm}

\begin{equation}
\label{eq:LoG3}
 {LoGSIM}[m,n]=\frac{ 2 \cdot LoG_1[m,n] \cdot LoG_2[m,n] +c_1} {(LoG_1[m,n])^2+(LoG_2[m,n])^2+c_2}
\end{equation}
 \vspace{-0.4cm}

Similarity metric becomes $1.0$ when the images are same and it gets closer to $0.0$ as the differences between images become very large. We set constants $c_1$ and $c_2$ to $0.001$ to avoid the issues when the denominator converges to $0.0$. 

\vspace{-0.2cm}

\subsection{Color Similarity}
\label{sec:main_Color}
\vspace{-0.2cm}
Color similarity is directly calculated over the \texttt{a} and the \texttt{b} channels separately. We use the similarity formulation as expressed in Eq. (\ref{eq:Col1}) and Eq. (\ref{eq:Col2}).

\vspace{-0.3cm}

\begin{equation}
\label{eq:Col1}
 {aSIM}[m,n]=\frac{ 2 \cdot a_1[m,n] \cdot a_2[m,n] +c_3} {(a_1[m,n])^2+(a_2[m,n])^2+c_4}
\end{equation}

\begin{equation}
\label{eq:Col2}
 {bSIM}[m,n]=\frac{ 2 \cdot b_1[m,n] \cdot b_2[m,n] +c_5} {(b_1[m,n])^2+(b_2[m,n])^2+c_6},
\end{equation}
%s
where $a_1[m,n]$ and $b_1[m,n]$ are the chroma channels in the reference image and  $a_2[m,n]$ and $b_2[m,n]$ are the chroma channels in the distorted image. $c_3$, $c_4$, $c_5$ and $c_6$ are the constants set to $0.001$.

\vspace{-0.3cm}
\subsection{Fusion}
\label{sec:main_Fusion}
\vspace{-0.2cm}
The human visual system is more sensitive to structural information compared to color. Based on this observation, chroma sub-sampling is introduced in image and video coding to assign less resolution to chroma information. $4$:$2$:$2$ is one of the most commonly used chroma sub-sampling format where chroma channels get half the resolution of luma channels. In the proposed work, we follow a similar approach and tune the significance of the intensity and the color-based components. The power of \texttt{LoG} similarity is set to $4.0$ and the powers of similarities for chroma channels are set to $2.0$. After this sensitivity adjustment, we choose the minimum among the similarity indexes as formulated in Eq. (\ref{eq:PerSIM2}) because the perceived quality is dominated by the most significant degradation.   

\vspace{-0.5cm}

\begin{multline}
\label{eq:PerSIM2}
LabSIM_{SR}[m,n]=\\
 min((LoGSIM[m,n])^4,(aSIM[m,n])^2,(bSIM[m,n])^2)
\end{multline}
 \vspace{-0.4cm}

We perform mean pooling to obtain a single quality value corresponding to the distortion map. Similarity is calculated over the full feature map so the pixels that are slightly distorted would bias the metric to be close to $1.0$. In order to increase the variation of the metric and spread the range of the estimations, we monotonically scale the resulting value with a power function  as given in  Eq. (\ref{eq:PerSIM1}).  

\vspace{-0.3cm}

\begin{equation}
\label{eq:PerSIM1}
PerSIM_{SR}=\left( \sum_{m=1}^{M} \sum_{n=1}^{N} \\
  \frac {LabSIM_{SR}[m,n]} {M \cdot N}
 \right) ^{c_7} 
\end{equation}

Mean pooling is performed over the whole image where $M$ is number of rows, $N$ is the number of columns and $c_7$ is the power index. $SR$ refers to single resolution since we use the reference and distorted images at the original resolution. Power index is set to $25$ so that the metric similarity index goes down to $0.0$ under severe degradation. Power indexes less than $25$ does not use the full metric range and indexes more than $25$ become extra sensitive to even slight degradation. This monotonic scaling does not bias the results since ranking-based validation metrics are insensitive to the monotonic mapping and the regression step before the linear correlation calculation perform monotonic mapping automatically.

\subsection{Multi-Resolution}
\label{sec:main_MultiRes}
\vspace{-0.2cm}
Perception in the visual system is hierarchical. At first, the raw data is acquired with the sensor-like structures. Then, the data is processed and transferred into different abstraction layers with varying resolutions. Different features and regions of interest can be more distinctive at different resolutions. Therefore, we calculate the perceptual similarity maps at different resolutions and fuse them together. We start by calculating \texttt{LoGSIM}, \texttt{aSIM} and \texttt{bSIM} over three different resolutions. The first set is calculated over the full resolution while the second and third are calculated at $3/5$ and $2/5$ times the full resolution, respectively. We refer to all the maps as $LoGSIM$, $aSIM$ and $bSIM$ and the scales of the resolution are shown with a subscript. \texttt{LoG} features and chroma similarities are extracted over the scaled maps and then interpolated to the original resolution using the bicubic approach. Since the average value and range of the metrics are not known, we directly calculate the geometric mean of the interpolated maps pixel-wise to obtain the multi-resolution perceptual similarity map as formulated in Eqs. (\ref{eq:PerSIM_MR1})-(\ref{eq:PerSIM_MR3}). 

\vspace{-0.7cm}

\begin{multline}
\label{eq:PerSIM_MR1}
LoGSIM_{MR}[m,n]\\
=\sqrt[3]{LoGSIM_{1.0} \cdot LoGSIM_{0.6} \cdot LoGSIM_{0.4}}
\end{multline}

\vspace{-0.8cm}

\begin{equation}
\label{eq:PerSIM_MR2}
aSIM_{MR}[m,n]=\sqrt[3]{aSIM_{1.0} \cdot aSIM_{0.6} \cdot aSIM_{0.4}}
\end{equation}
\begin{equation}
\label{eq:PerSIM_MR3}
bSIM_{MR}[m,n]=\sqrt[3]{bSIM_{1.0} \cdot bSIM_{0.6} \cdot bSIM_{0.4}}
\end{equation}

Multi-resolution indexes are combined in the same way as the single resolution given in Eq. (\ref{eq:PerSIM_MR4}).

\vspace{-0.7cm}
\begin{multline}
\label{eq:PerSIM_MR4}
LabSIM_{MR}[m,n]= min((LoGSIM_{MR}[m,n])^4,\\
(aSIM_{MR}[m,n])^2,(bSIM_{MR}[m,n])^2) 
\end{multline}

Finally, multi-resolution perceptual quality map is mean pooled and monotonically mapped as formulated in Eq. (\ref{eq:PerSIM1_MR5}).

\vspace{-0.5cm}

\begin{equation}
\label{eq:PerSIM1_MR5}
PerSIM=\left(\sum_{m=1}^{M} \sum_{n=1}^{N} \\
  \frac {LabSIM_{MR}[m,n]} {M \cdot N}
 \right )^{25} 
\end{equation}
\vspace{-0.4cm}

As the resolution gets lower, it becomes more challenging to detect distinctive features. Therefore, we decrease the block-size and standard deviation accordingly as tabulated in Table \ref{tab:PerSIM_MR}. The scale values, standard deviation and block size are selected by visually assessing the distinctiveness of randomly selected feature maps.  

\vspace{-0.2cm}

\begin{table}[htbp!]
\vspace{-0.3cm}

  \caption{Multi-resolution PerSIM parameters}
%    \vspace{-0.2cm}

  \centering
    \footnotesize
  
    \begin{tabular}{ccc}   \hline

	\textbf{Scaling Ratio} &\textbf{Standard Deviation} &\textbf{Block Size}\\  \hline

	1.0 &10.0 &13x13\\  \hline
    0.6 &8.0 &4x4\\  \hline
	0.4 &7.0 &2x2\\  \hline

    \end{tabular}%
  \label{tab:PerSIM_MR}
  \vspace{-0.1cm}

\end{table}

%
%lSUM=nthroot(lSIM.*lSIM_0_6.*lSIM_0_4,3);
%aSUM=nthroot(aSIM.*aSIM_0_6.*aSIM_0_4,3);
%bSUM=nthroot(bSIM.*bSIM_0_6.*bSIM_0_4,3);

%outCT=mean2(min(lSUM.^4,min(aSUM.^2,bSUM.^2)))^25;

\begin{table}[htbp!]

  \caption{LIVE Results}
  \label{tab:LIVE_Results}
%\vspace{-0.4cm}

  \centering

    \footnotesize
  
    \begin{tabular}{c||cccccc}   \hline

    \multirow{2}[3]{*}{\textbf{Sequence}}

          & \textbf{Jp2k}& \textbf{Jpeg}& \textbf{Wn}& \textbf{Gblur}& \textbf{FF}  & \textbf{All}   
          
 \\ \cline{2-7}

    & \multicolumn{6}{c}{\textbf{Pearson (PLCC)}}               \\ \hline

	\textbf{PSNR} &0.923 &0.913 &0.945 &0.843 &0.887 &0.898\\

  \textbf{SSIM} &\textbf{0.963} &0.957 &0.976 &0.940 &\textbf{0.956} &0.945 \\

  \textbf{MS-SSIM} &0.962 &\textbf{0.961} &\textbf{0.977} &0.943 &\textbf{0.948} &0.946 \\

 \textbf{IW-SSIM} &0.959 &\textbf{0.959} &\textbf{0.981} &\textbf{0.957} &\textbf{0.953} &\textbf{0.951}  \\

 \textbf{FSIMc} &0.960&0.953&\textbf{0.977}&\textbf{0.955}&\textbf{0.953}&0.950 \\ 

 \textbf{PSNR-HA} &\textbf{0.976}&\textbf{0.971}&0.980&0.935&\textbf{0.953}&\textbf{0.953}
 \\

 \textbf{CW-SSIM} &0.926 &0.927 &0.949 &0.768 &0.835 &0.872  \\ 
 
  \textbf{LogSIM} &0.956 &0.952 &\textbf{0.987} &0.943 &0.942 &0.943  \\ \hline

	\textbf{PerSIM}  &\textbf{0.976}&\textbf{0.959}&0.968&\textbf{0.967}&0.946&\textbf{0.955}   \\ \hline

    & \multicolumn{6}{c}{\textbf{RMSE}}           \\ \hline    
 
 	\textbf{PSNR} &9.92&10.10&8.34&11.80&10.22&10.12 \\  
	
	  \textbf{SSIM} &\textbf{7.11}&7.74&8.65&7.54&6.45&7.52 \\  

  \textbf{MS-SSIM} &7.12&\textbf{7.30}&8.38&7.38&\textbf{7.04}&7.43 \\  

 \textbf{IW-SSIM} &7.38&\textbf{7.64}&\textbf{6.95}&\textbf{6.37}&\textbf{6.86}&\textbf{7.11}  \\ 

 \textbf{FSIMc} &7.55&7.73&6.97&\textbf{6.71}&\textbf{6.68}&\textbf{7.20}  \\ 

 \textbf{PSNRHA}  &\textbf{6.59}&\textbf{6.89}&\textbf{5.98}&8.27&6.83&6.93  \\

 \textbf{CW-SSIM} &9.75&9.30&9.24&14.45&13.62&10.87 \\

 \textbf{LogSIM} &7.93&8.36&\textbf{6.06}&7.70&7.71&7.66 \\ \hline

 \textbf{PerSIM} &\textbf{6.22}&\textbf{7.30}&7.25&\textbf{5.68}&7.36&\textbf{6.80} \\ \hline
	
%	\textbf{LoG}  &0.976 &0.957 &\textbf{0.989}  &0.950  &0.915 &0.947 &0.874 &0.830 &\textbf{0.914} &0.813  &0.774  &0.813  \\ \hline

%	\textbf{BSD}  &0.942 &0.931 &0.962  &0.954  &0.967 &0.922 &0.797 &0.784 &0.837 &0.830  &0.855  &0.767  \\ \hline

    \end{tabular}%
    \vspace{-0.4cm}

\end{table}

\vspace{-0.7cm}

\section{Validation}
\label{sec:validation}

  \vspace{-0.3cm}

LIVE and TID2013 image databases are used in the validation of \texttt{PerSIM}. LIVE database includes 29 reference images and 779 degraded images under the distortion of  JPEG, JPEG2000 (Jp2k), White Noise (Wn), Gaussian blur (Gblur) and Fast Fading Rayleigh channel errors (FF). TID2013 consists of 25 reference images that are originally from Kodak Lossless True Color Image Suite \cite{kodak}. Reference images are degraded with 24 different types of distortions that fall into the categories of Noise, Actual, Simple, Exotic, New and Color. TID2013 database is introduced in \cite{TID2013} where ranking based metrics Spearman and Kendall correlation coefficients are used for the validation. Therefore, we follow the same approach as in \cite{TID2013}.

Objective image quality metrics are defined in different numerical ranges and monotonic regression is necessary for a fair comparison if validation includes metrics that are based on linearity and accuracy including but not limited to Pearson linear correlation coefficient (PLCC) and root mean-squared error (RMSE). In the literature, validation of the metrics in the LIVE database are mostly based on PLCC and RMSE and the function formulated in Eq. (\ref{eq:nonlinreg}) is used for monotonic regression. Therefore, we also calculate the PLCC and the RMSE after the monotonic regression as tabulated in Table \ref{tab:LIVE_Results}.

\vspace{-0.4cm}

\begin{equation}
\label{eq:nonlinreg}
S=\beta_1 \left ( \frac{1}{1}-\frac{1}{2+exp(\beta_2(S_0 -\beta_3 ))} \right )+\beta_4 S_0 +\beta_5
\end{equation}

\vspace{-0.2cm}

We use the TID2013 benchmark data to compare the proposed metric with the state of the art. In the LIVE database, we use the common structural metrics and the best perfoming ones in the TID benchmark. In order to show the effect of color similarity in the overall metric, we replace the $LabSIM_{MR}[m,n]$ in Eq. \ref{eq:PerSIM1_MR5} with $LoGSIM_{MR}[m,n]$ and report the results as $LogSIM$. Top three performance values are highlighted in the results tables to indicate best performing metrics.  In the LIVE database, \texttt{PerSIM} is among the top metrics in the compression-related degradation JPEG2000 and JPEG and also in Gaussian blur. However, \texttt{PerSIM} is not as good as structure and phase conjugacy-based metrics in case of White noise and Fastfaing artifacts.  White noise artifacts are captured by the \texttt{LoG} features but color similarity is less sensitive to these artifacts. In case of the Fastfading, communication channel errors can lead to local errors that are perceptually very disturbing but they would be overlooked by \texttt{PerSIM} since relative size of the errors can be negligible compared to the rest of the sharp transitions in the image. In the overall LIVE database, \texttt{PerSIM} still performs better than the compared metrics.

\begin{table*}[htbp!]
  \label{tab:TID_Results}
    \caption{TID2013 Results}

  \centering
    \footnotesize
  
    \begin{tabular}{c||ccccccc||ccccccc}   \hline

%    \multirow{2}[3]{*}{Sequence} & \multicolumn{7}{|c|}{Spearman}  \\ %\cline{2-8}
%          & Noise& Actual& Simple& Exotic& New  & Color &Full  \\ \hline

    \multirow{2}[3]{*}{\textbf{Sequence}} 
    
    & \multicolumn{7}{c||}{\textbf{Spearman (SROCC)}}     & \multicolumn{7}{c}{\textbf{Kendall (KCC)}}  \\ \cline{2-15}

          & \textbf{Noise}& \textbf{Actual}& \textbf{Simple}& \textbf{Exotic}& \textbf{New}  & \textbf{Color} &\textbf{Full}  
          & \textbf{Noise}& \textbf{Actual}& \textbf{Simple}& \textbf{Exotic}& \textbf{New}  & \textbf{Color} &\textbf{Full}

          \\ \hline

	\textbf{FSIM-c} &0.902 &0.915 &0.947 &\textbf{0.841} &\textbf{0.788}  &\textbf{0.775}  &\textbf{0.851}  &0.722 &0.742 &0.792 &\textbf{0.651} &\textbf{0.611}  &\textbf{0.592}  &\textbf{0.666}\\ 
    \textbf{PSNR-HA} &\textbf{0.923} &\textbf{0.938} &\textbf{0.953} &0.825 &0.701  &0.632  &\textbf{0.819}  &\textbf{0.760} &\textbf{0.787} &\textbf{0.818} &0.624 &0.541  &0.477  &\textbf{0.643}\\ 
    \textbf{PSNR-HMA} &0.915 &\textbf{0.934} &0.937 &0.814 &0.738  &0.675  &0.813 &\textbf{0.745} &\textbf{0.777} &0.785 &0.610 &0.572  &0.507  &0.631 \\ 
	\textbf{FSIM} &0.897 &0.911 &0.949 &\textbf{0.844} &0.649  &0.565  &0.801  &0.715 &0.736 &0.795 &\textbf{0.655} &0.518  &0.447  &0.629\\ 
	\textbf{MS-SSIM} &0.873 &0.887 &0.905 &\textbf{0.841} &0.631  &0.566  &0.787  &0.679 &0.697 &0.720 &\textbf{0.647} &0.490  &0.450  &0.607\\  
	\textbf{IW-SSIM} &0.871 &0.887 &0.911 &0.840 &0.619 &0.549 &0.778 &0.678 &0.701 &0.730 &0.644 &0.475 &0.424 &0.597\\ 
	\textbf{PSNRc} &0.769 &0.803 &0.876 &0.562 &\textbf{0.777}  &\textbf{0.734}  &0.687  &0.562 &0.596 &0.689 &0.392 &\textbf{0.576}  &\textbf{0.536}  &0.496\\ 
	\textbf{VSNR} &0.869 &0.882 &0.912 &0.706 &0.589  &0.512  &0.681  &0.676 &0.690 &0.731 &0.519 &0.437  &0.378  &0.508\\ 	
	\textbf{PSNR-HVS} &\textbf{0.917} &0.926 &\textbf{0.951} &0.601 &0.646  &0.555  &0.654  &\textbf{0.754} &0.766 &\textbf{0.809} &0.435 &0.512  &0.441  &0.507\\ 
	\textbf{PSNR} &0.822 &0.825 &0.913 &0.597 &0.618  &0.535  &0.640  &0.623 &0.624 &0.745 &0.425 &0.468  &0.408  &0.470\\ 
	\textbf{SSIM} &0.757 &0.788 &0.837 &0.632 &0.579  &0.505  &0.637  &0.551 &0.577 &0.628 &0.455 &0.418  &0.378  &0.463\\ 	
	\textbf{NQM} &0.836 &0.857 &0.875 &0.589 &0.625  &0.538  &0.635  &0.641 &0.666 &0.681 &0.412 &0.478  &0.401  &0.466\\ 
	\textbf{PSNR-HVS-M} &0.906 &0.917 &0.938 &0.564 &0.646  &0.553  &0.625  &0.733 &0.749 &0.780 &0.403 &0.513  &0.433  &0.481\\ 
	\textbf{VIFP} &0.784 &0.815 &0.897 &0.557 &0.589  &0.506  &0.608  &0.587 &0.621 &0.714 &0.406 &0.445  &0.385  &0.456\\ 
	\textbf{WSNR} &0.880 &0.897 &0.933 &0.423 &0.646  &0.555  &0.580  &0.696 &0.718 &0.772 &0.297 &0.510  &0.429  &0.446\\ 
	\textbf{LogSIM} &0.910&0.923&0.947 &0.806&0.662&0.604&0.787&0.736&0.756&0.799&0.615&0.521
	&0.473 &0.618  \\ \hline	
	\textbf{PerSIM} &\textbf{0.925}&\textbf{0.936}&\textbf{0.950} &0.799&\textbf{0.863}&\textbf{0.856}&\textbf{0.854}&\textbf{0.760}&\textbf{0.778}&\textbf{0.807}&0.606&\textbf{0.681}
	&\textbf{0.674} &\textbf{0.677}  \\ \hline

    \end{tabular}%
    \vspace{-2.5mm}

\vspace{-0.28 cm}

\end{table*}

\begin{figure}[h!]

\begin{minipage}[b]{0.46\linewidth}

  \centering
\includegraphics[width=0.85\linewidth, trim= 20mm 72mm 20mm 65mm]{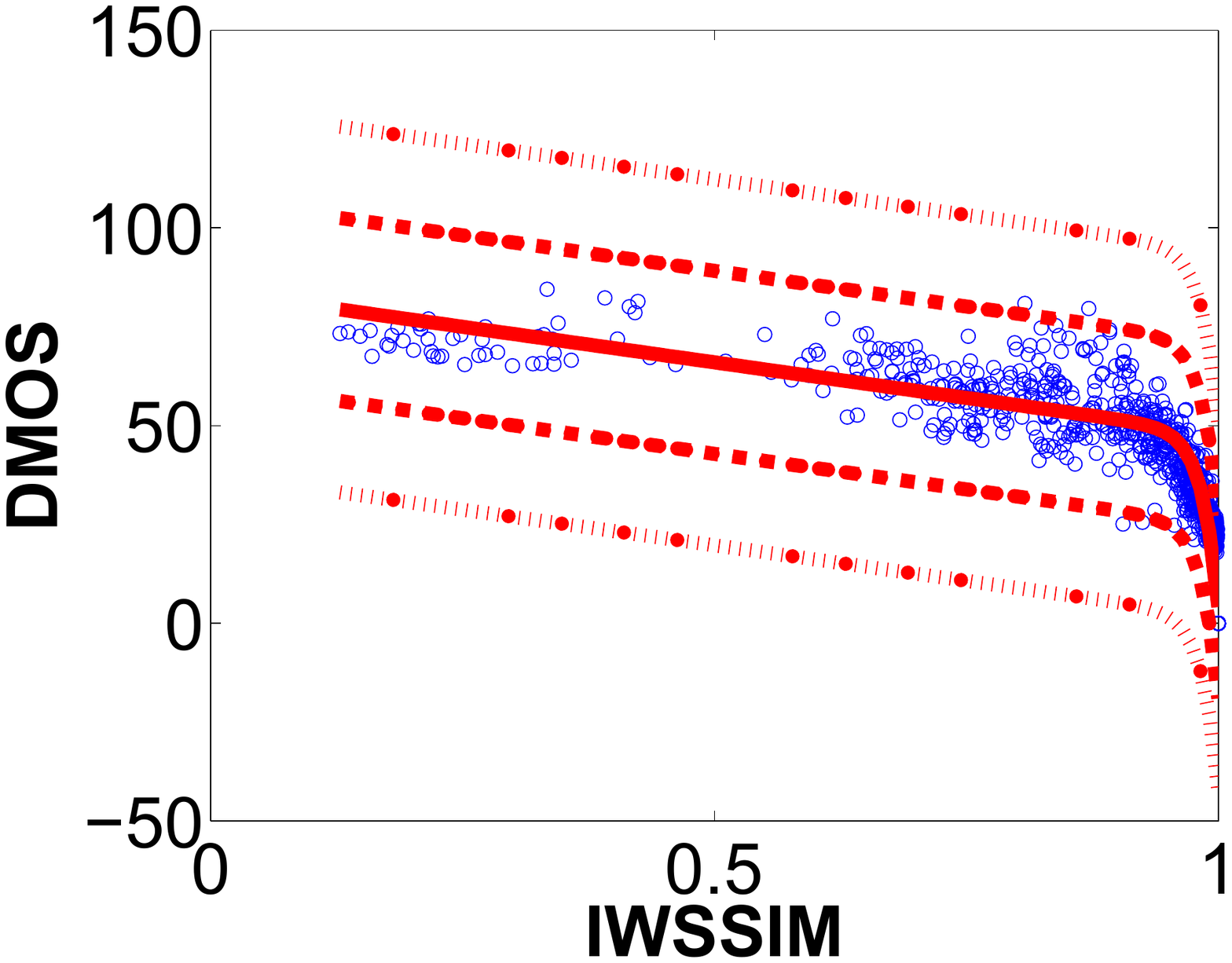}
  \vspace{-0.33 cm}
  \centerline{\footnotesize{(a) LIVE-IW-SSIM   } }%\medskip
\end{minipage}
  \vspace{0.20cm}
\hfill
\begin{minipage}[b]{0.46\linewidth}
  \centering
\includegraphics[width=0.80\linewidth, trim= 20mm 72mm 20mm 65mm]{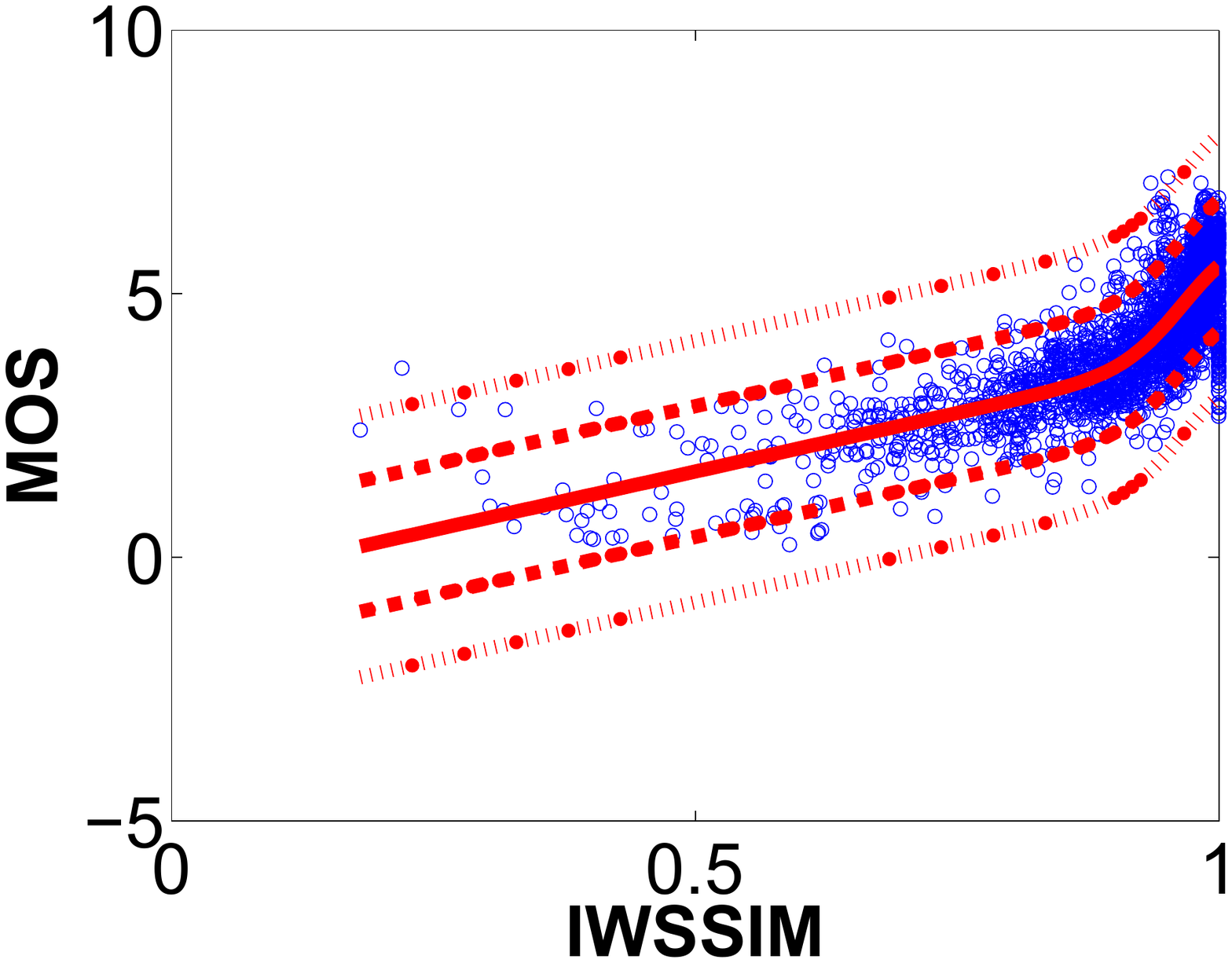}
  \vspace{-0.33 cm}
  \centerline{\footnotesize{(b) TID-IW-SSIM }}%\medskip
\end{minipage}
\begin{minipage}[b]{0.46\linewidth}
  \centering
\includegraphics[width=0.80\linewidth, trim= 20mm 72mm 20mm 65mm]{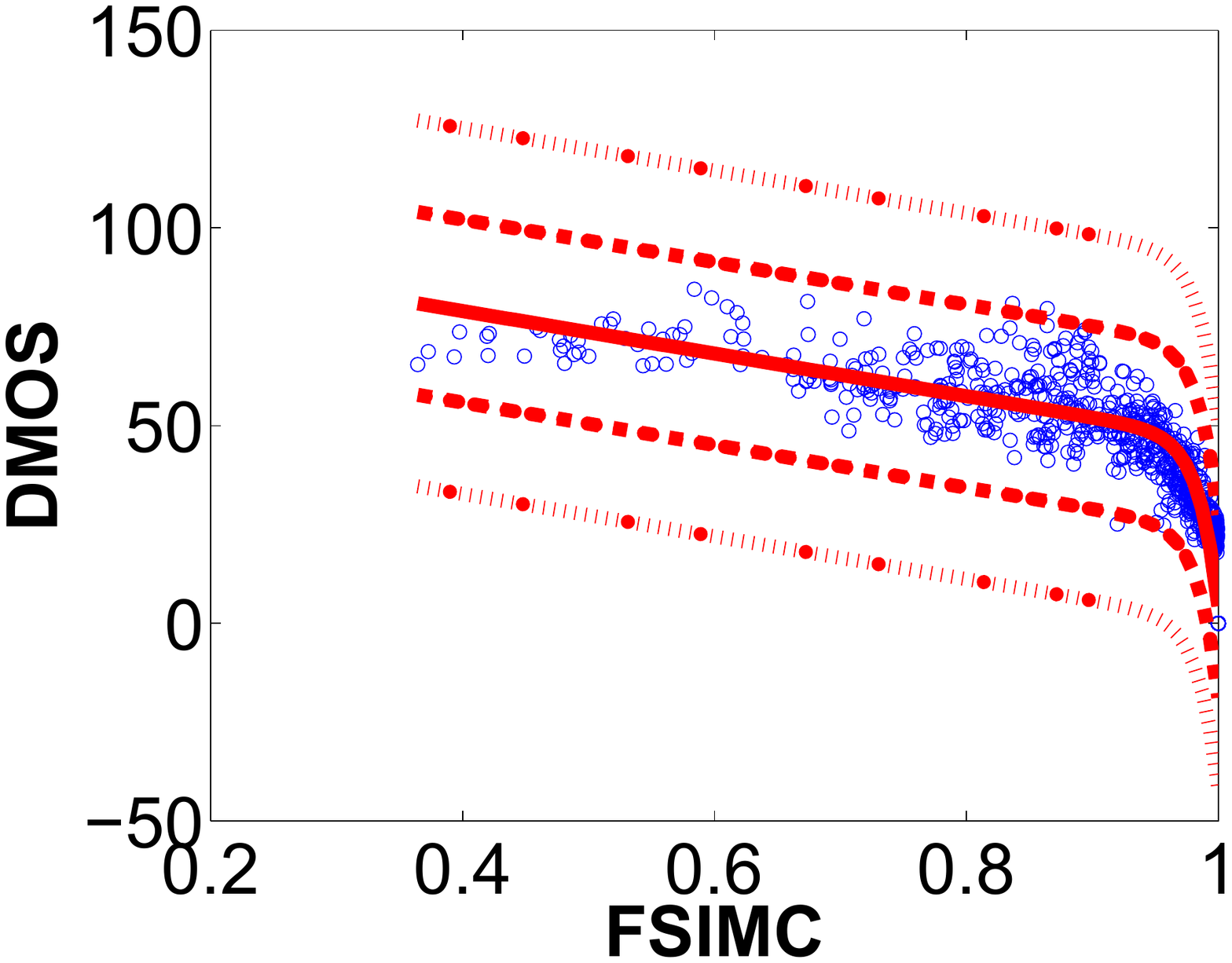}
  \vspace{-0.33 cm}
  \centerline{\footnotesize{(c)  LIVE-FSIM-c  } }%\medskip
\end{minipage}
  \vspace{0.2cm}
\hfill
\begin{minipage}[b]{0.46\linewidth}
  \centering
\includegraphics[width=0.80\linewidth, trim= 20mm 72mm 20mm 65mm]{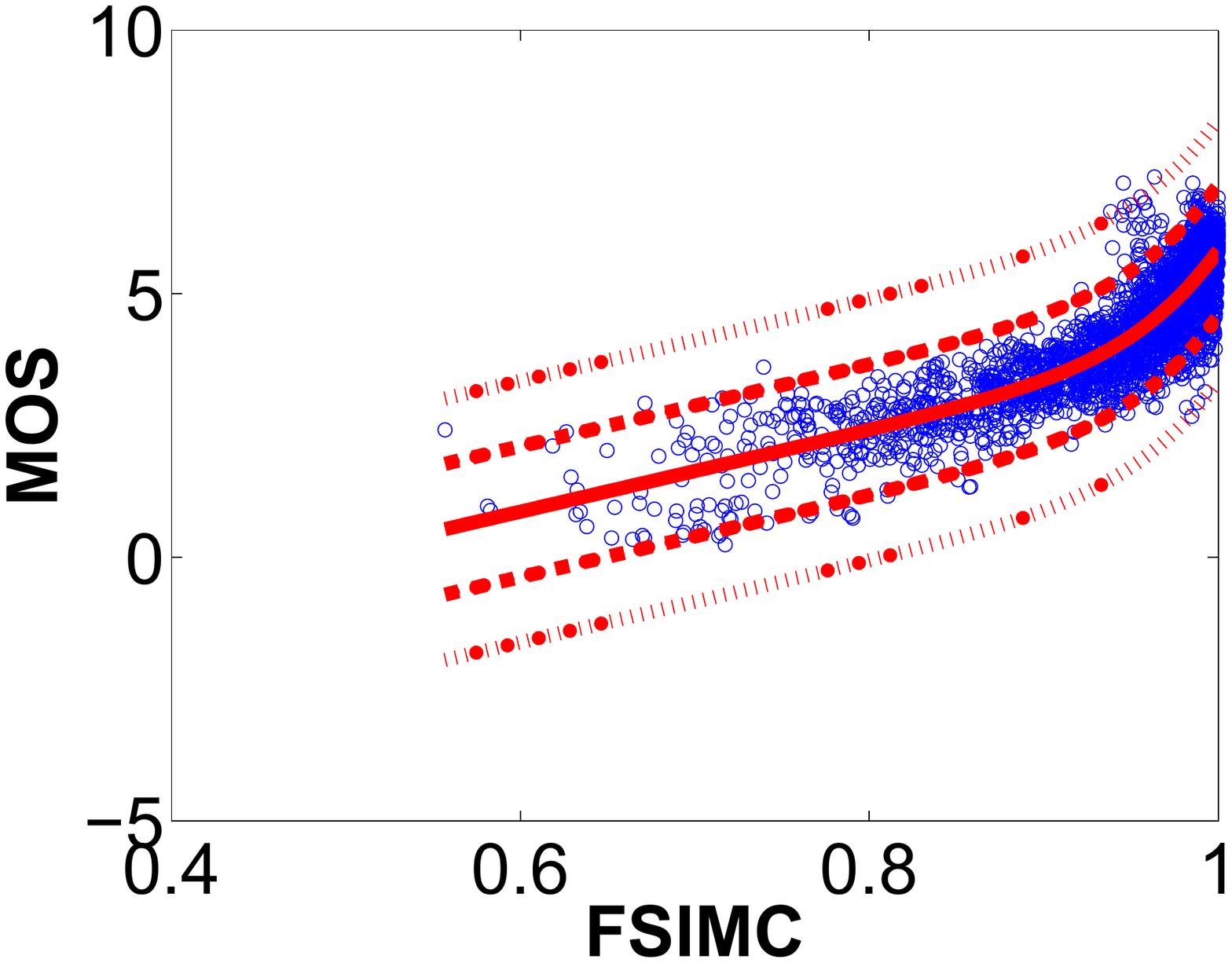}
  \vspace{-0.33 cm}
  \centerline{\footnotesize{(d) TID-FSIM-c}}%\medskip
\end{minipage}
\begin{minipage}[b]{0.46\linewidth}
  \centering
\includegraphics[width=0.80\linewidth, trim= 20mm 72mm 20mm 65mm]{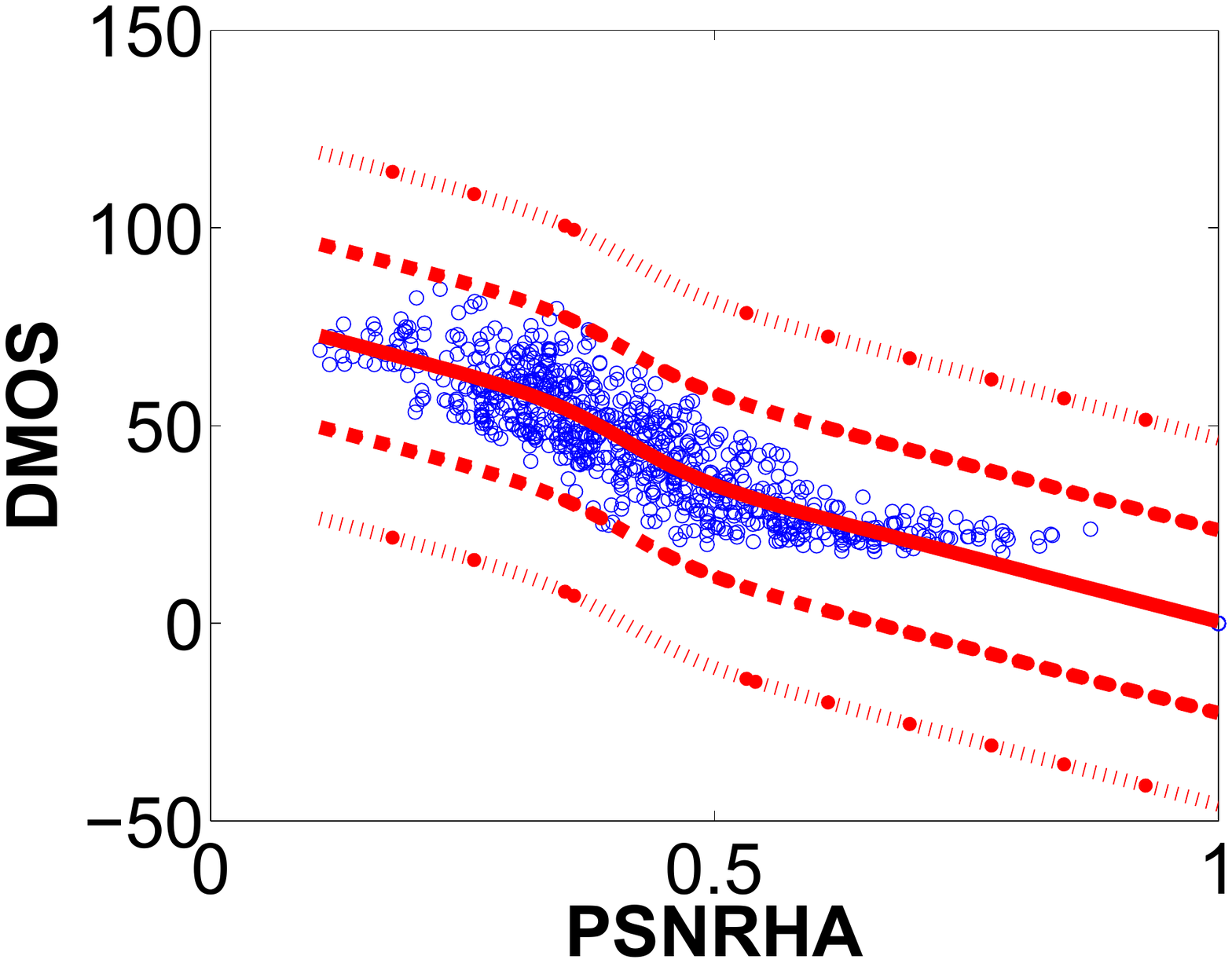}
  \vspace{-0.33 cm}
  \centerline{\footnotesize{(e) LIVE-PSNR-HA  } }%\medskip
\end{minipage}
  \vspace{0.2cm}
\hfill
\begin{minipage}[b]{0.46\linewidth}
  \centering
\includegraphics[width=0.80\linewidth, trim= 20mm 72mm 20mm 65mm]{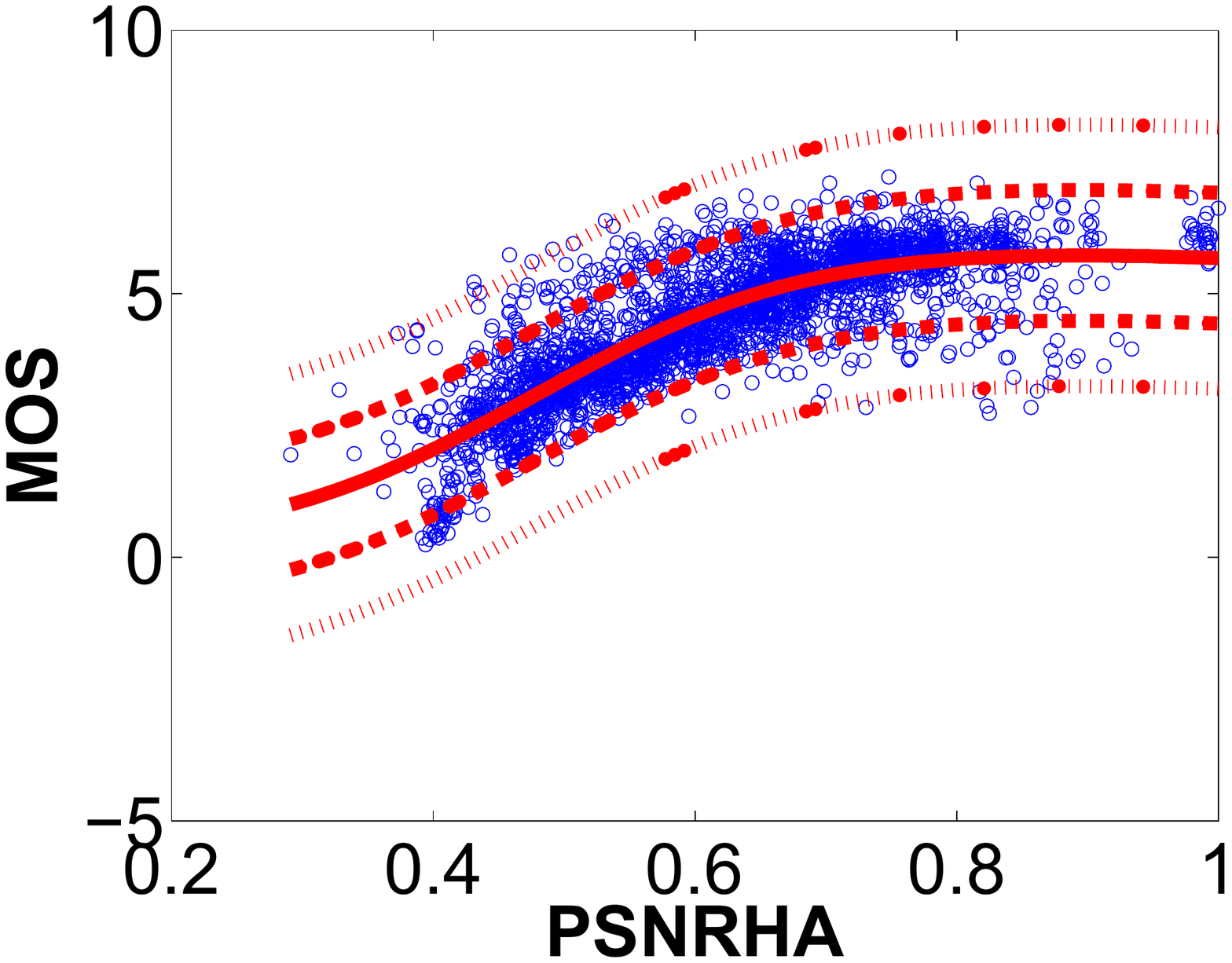}
  \vspace{-0.33 cm}
  \centerline{\footnotesize{(f) TID-PSNR-HA}}%\medskip
\end{minipage}
\begin{minipage}[b]{0.46\linewidth}
  \centering
\includegraphics[width=0.80\linewidth, trim= 20mm 72mm 20mm 65mm]{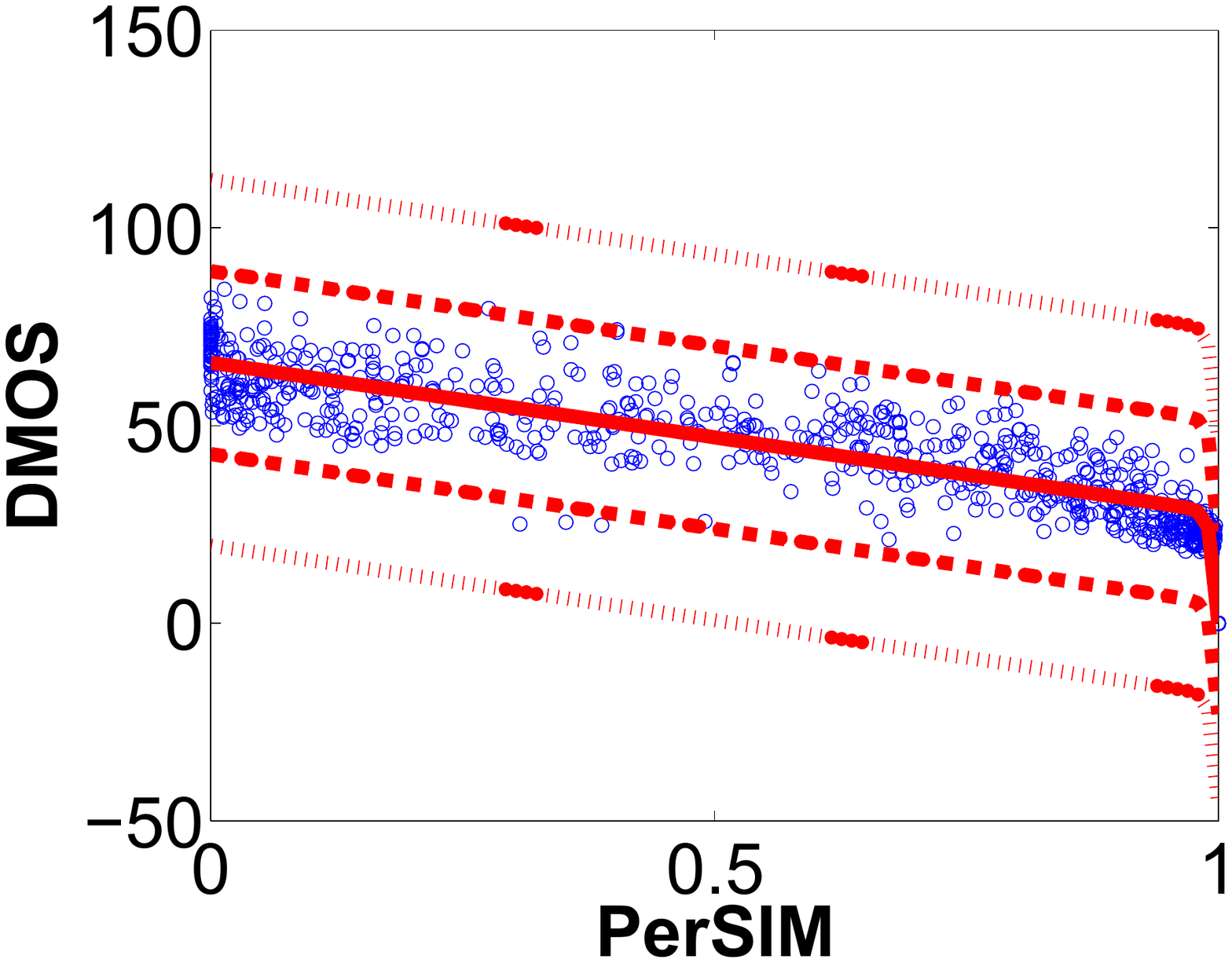}
  \vspace{-0.33 cm}
  \centerline{\footnotesize{(g)  LIVE-PerSIM } }%\medskip
\end{minipage}
  \vspace{0.2cm}
\hfill
\begin{minipage}[b]{0.46\linewidth}
  \centering
\includegraphics[width=0.80\linewidth, trim= 20mm 72mm 20mm 65mm]{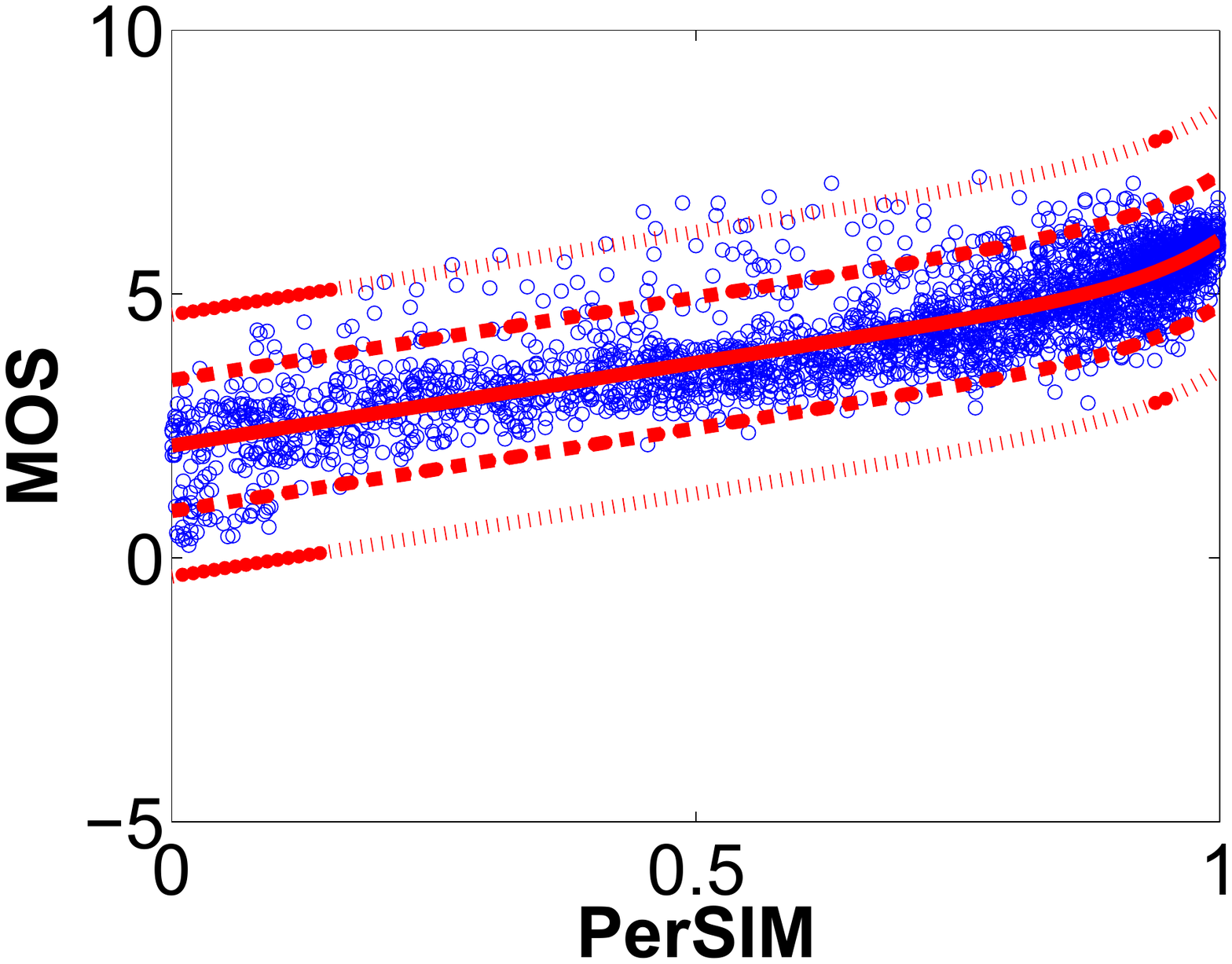}
  \vspace{-0.40 cm}
  \centerline{\footnotesize{(h)TID-PerSIM}}%\medskip
\end{minipage}
  \vspace{-0.2 cm}
\caption{Scatter plots of objective image quality metrics values}
 % \vspace{-0.20 cm}

\label{fig:Scatter}
\vspace{-0.60 cm}
\end{figure}

The performance of \texttt{PerSIM} in the TID2013 databse is tabulated in Table \ref{tab:TID_Results}. According to the validation results, \texttt{PerSIM} is among the best performing metrics in all the categories except Exotic. Exotic category includes local block-wise distortions and JPEG2000 tranmission errors that can lead to local and slice-based distortions that are overlooked by the \texttt{LoG} features and color similarity. In the overall TID2013, \texttt{PerSIM} is still the best performing objective quality metric among the compared ones. As it can be seen in Table \ref{tab:LIVE_Results} and Table \ref{tab:TID_Results}, the performance of the quality metric over the full image set  degrades without the color similarity in both LIVE and TID2013.

If we compare the metrics in both LIVE and TID2013 databases, the best performing metrics are IW-SSIM, FSIMc, PSNR-HA and PerSIM. The scatter plots of the best performing metrics are given in  Fig. \ref{fig:Scatter} to observe the distributional characteristics. Information theoretic-weighting based structural metric (IW-SSIM) scores mostly cluster around the high quality region and the same observation is valid for the metric based on phase-conjugacy (FSIMc). PSNR-HA has an outlier problem when we include identical images and the range of the metric is not bounded. In order to obtain the scatter plot in  Fig. \ref{fig:Scatter}(f), we exclude the outliers. PSNR-HA estimates are mostly centered in the metric range and it has a higher linearity compared to the strucutral and phase-conjugacy-based metrics. \texttt{PerSIM} scores are distributed in the full metric range and show a high linearity. Almost all the estimates are in the one standard deviation range in the LIVE database. However, in the TID2013 database, some of the estimates are located between one and two standard deviation and only a minority is located outside of two standard deviation range. Most of the outliers in the TID2013 database correspond to the Exotic class since \texttt{PerSIM} has difficulty in capturing local degradations.

\vspace{-0.55cm}

\section{Conclusion}
\label{subsec:conc}
\vspace{-0.40cm}

We proposed a full reference multi-resolution image quality metric based on \texttt{LoG} features and chroma similarity in the perceptually uniform \texttt{Lab} domain. \texttt{LoG} features are used to model the retinal ganglion cells in the human visual system and the color similarity complements the structural similarity. The results in the LIVE and TID2013 database show that \texttt{PerSIM} outprforms state of the art metrics in terms of monotonicity, accuracy and linearity. Even \texttt{PerSIM} detects  majority of the distortions accurately, it overlooks local distortions. As an ongoing work, we are working on a smarter pooling strategy to make the metric sensitive to local distortions.

\newpage

% -------------------------------------------------------------------------
%\bibliographystyle{IEEEbib}
%\bibliography{refs}

\begin{thebibliography}{10}

\bibitem{Egiazarian}
K.~Egiazarian, J.~Astola, N.~Ponomarenko, V.~Lukin, F.~Battisti, and M.~Carli,
\newblock ``{A New Full-reference Quality Metrics based on HVS},''
\newblock in {\em Proceedings of the Second International Workshop on Video
  Processing and Quality Metrics,}, 2006.

\bibitem{Ponomarenko2007}
N.~Ponomarenko, F.~Silvestri, K.~Egiazarian, M.~Carli, J.~Astola, and V.~Lukin,
\newblock ``On between-coefficient contrast masking of dct basis functions,''
\newblock in {\em Proceedings of the Second International Workshop on Video
  Processing and Quality Metrics}, 2007, pp. 1--4.

\bibitem{Ponomarenko2011}
N.~Ponomarenko, O.~Ieremeiev, V.~Lukin, K.~Egiazarian, and M.~Carli,
\newblock ``{Modified Image Visual Quality Metrics for Contrast Change and Mean
  Shift Accounting},''
\newblock {\em Proceedings of CADSM}, 2011.

\bibitem{Mitsa1993}
T.~Mitsa and K.~L. Varkur,
\newblock ``{Evaluation of Contrast Sensitivity Functions for the Formulation
  of Quality Measures Incorporated in Halftoning Algorithms},''
\newblock in {\em ICASSP}, 1993.

\bibitem{Chandler2007}
D.~M. Chandler and S.~S. Hemami,
\newblock ``{VSNR: a wavelet-based visual signal-to-noise ratio for natural
  images.},''
\newblock {\em IEEE Transactions on Image Processing}, vol. 16, no. 9, pp.
  2284--98, Sept. 2007.

\bibitem{Wang2004}
Z.~Wang, A.~C. Bovik, H.~R. Sheikh, and E.~P. Simoncelli,
\newblock ``{Image quality assessment: from error visibility to structural
  similarity.},''
\newblock {\em IEEE transactions on image processing : a publication of the
  IEEE Signal Processing Society}, vol. 13, no. 4, pp. 600--12, Apr. 2004.

\bibitem{Wang2003}
Z.~Wang, E.~P. Simoncelli, and A.~C. Bovik,
\newblock ``{Multi-Scale Structural Similarity For Image Quality Assessment (
  Invited Paper )},''
\newblock {\em the Thirty-Seventh Asilomar Conference on Signals, Systems and
  Computers}, vol. 2, pp. 9--13, 2004.

\bibitem{Inter2005}
Z.~Wang and E.~P. Simoncelli,
\newblock ``{Translation Insensitive Image Similiarity In Complex Wavelet
  Domain Zhou Wang and Eero P . Simoncelli},''
\newblock vol. II, no. March, pp. 573--576, 2005.

\bibitem{Wang2011}
Z.~Wang and Q.~Li,
\newblock ``{Information Content Weighting for Perceptual Image Quality
  Assessment.},''
\newblock {\em IEEE Transactions on Image Processing}, vol. 20, no. 5, pp.
  1185--98, May 2011.

\bibitem{Zhang2011}
L.~Zhang, L.~Zhang, X.~Mou, and D.~Zhang,
\newblock ``{FSIM: A Feature Similarity Index for Image Quality Assessment.},''
\newblock {\em IEEE Transactions on Image Processing}, vol. 20, no. 8, pp.
  2378--86, Aug. 2011.

\bibitem{Feng2014}
W.~Xue, X.~Mou, L.~Zhang, A.C. Bovik, and X.~Feng,
\newblock ``{Blind Image Quality Assessment Using Joint Statistics of Gradient
  Magnitude and Laplacian Features},''
\newblock {\em Image Processing, IEEE Transactions on}, vol. 23, no. 11, pp.
  4850--4862, Nov 2014.

\bibitem{log2014}
X.~Mou, W.~Xue, C.~Chen, and L.~Zhang,
\newblock ``{LoG Acts as a Good Feature in the Task of Image Quality Assessment
  },''
\newblock {\em Proc. SPIE}, vol. 9023, pp. 902313--902313--7, 2014.

\bibitem{Robson1966}
C.~Enroth-Cugell and J.~G. Robson,
\newblock ``{The Contrast Sensitivity of Retinal Ganglion Cells of the Cat},''
\newblock {\em The Journal of Physiology}, 1966.

\bibitem{Young1987}
R.~A. Young,
\newblock ``{The Gaussian Derivative Model for Spatial Vision: I. Retinal
  Mechanisms},''
\newblock {\em Spatial Vision}, 1987.

\bibitem{kodak}
Eastman~Kodak Company,
\newblock ``{Lossless True Color Image Suite},'' http://r0k.us/graphics/kodak/,
\newblock [Online].

\bibitem{TID2013}
N.~Ponomarenko, O.~Ieremeiev, V.~Lukin, K.~Egiazarian, L.~Jin, J.~Astola,
  B.~Vozel, K.~Chehdi, M.~Carli, F.~Battisti, and C.-C.J. Kuo,
\newblock ``{Color Image Database TID2013: Peculiarities and Preliminary
  Results},''
\newblock pp. 106--111, June 2013.

\end{thebibliography}

\end{document}